\documentclass[doublecol]{epl2} 

\usepackage[utf8]{inputenc}  
\usepackage[T1]{fontenc} 

\title{Nonlinear scattering of atomic bright solitons in disorder}
\shorttitle{Nonlinear scattering of atomic bright solitons in disorder} 

\author{A. Boissé \and G. Berthet \and L. Fouch\'e \and G. Salomon \and A. Aspect \and S. Lepoutre \and T. Bourdel \inst{1}}

\shortauthor{A. Boissé \etal}
\institute{                    
  \inst{1} Laboratoire Charles Fabry, Institut d'Optique, CNRS, Univ Paris-Sud - 2, Avenue Augustin Fresnel, 91127 PALAISEAU CEDEX, France
}


\abstract{We observe nonlinear scattering of  $^{39}$K atomic bright solitons launched in a one-dimensional (1D) speckle disorder. We directly compare it with the scattering of non-interacting particles in the same disorder. The atoms in the soliton tend to be collectively  either reflected or transmitted, in contrast with the behavior of independent particles in the single scattering regime, thus demonstrating a clear nonlinear effect in scattering. The observed strong fluctuations in the reflected fraction, between zero and 100\%, are interpreted as a consequence of the strong sensitivity of the system to the experimental conditions and in particular to the soliton velocity. This behavior is reproduced in a mean-field framework by Gross-Pitaevskii simulations, and mesoscopic quantum superpositions of the soliton being fully reflected and fully transmitted are not expected for our parameters. We discuss the conditions for observing such superpositions, which would find applications in atom interferometry beyond the standard quantum limit. 
}

\begin{document}

\maketitle

\section{Introduction}

The physics of transport of particles in disorder is associated with different scenarios. In absence of interaction, the simplest description is based on diffusion \cite{Robert2010}, but  the coherence of the matter waves describing the particles can play a role, as in the phenomena  of coherent backscattering \cite{Cherroret2012, Jendrzejewski2012a, Muller2015} and  Anderson localization  \cite{Anderson1958, VanTiggelen1999,  Sanchez-Palencia2007, Chabe2008, Billy2008, Roati2008, Lagendijk2009, Kondov2011, Jendrzejewski2012, Manai2015}.  However, in many physical systems, interactions cannot be ignored. In condensed matter physics, interactions between electrons can strongly affect electric conductivity \cite{Lee1985} and in optics, high intensity light induces a nonlinear response of dielectrics, leading for instance to the optical Kerr effect, and thus spatial and/or temporal fluctuations of the index of refraction. Understanding the interplay between disorder and interactions in the transport of quantum particles  is thus an important challenge. 

In a mean-field approach, one can use nonlinear wave equations in disordered media \cite{Flach2014, Cherroret2014} in order to describe experimental observations of the competition between a weak nonlinearity and localization, in optics \cite{Schwartz2007, Lahini2008} or in ultra-cold quantum gases  \cite{Lucioni2011}. Beyond the mean-field approximation, many-body localization phenomena, leading to non-ergodic behavior, are predicted \cite{Giamarchi1988, Basko2006}. In this context, several problems of transport of interacting quantum gases in disorder have been studied \cite{Fallani2007, Chen2008, White2009, Deissler2010, Pasienski2010, Gadway2011, Allard2012, Kondov2015, Schreiber2015, Meldgin2016, Choi2016}. We report here on a new phenomenon of nonlinear  transport of quantum particles: nonlinear scattering of atomic bright solitons in an optical disorder.

A soliton is a stable non-spreading wave-packet, solution of a nonlinear wave equation, where a strong nonlinearity compensates dispersion. Solitons  are ubiquitous in nonlinear wave physics \cite{Malomed2005, Dauxois2006}. Their propagation in a disordered medium is intriguing since the effect of the nonlinearity cannot  be treated as a small perturbation of the non interacting problem \cite{Kivshar1990}. An atomic  bright soliton is a 1D Bose-Einstein condensate of atoms with attractive interactions \cite{Khaykovich2002, Strecker2002}. At the mean field level, it is described by the Gross-Pitaevskii equation, which is identical to the so called  "nonlinear Schr\"odinger equation" used to describe the 1D propagation of light in Kerr media. This approach has been used to numerically study soliton non-linear scattering on a narrow barrier \cite{Lee2006, Hansen2012, Gertjerenken2012, Helm2012, Cuevas2013, Helm2014}. Qualitatively similar results for a 1D disordered potential in the single scattering regime can be expected. Experimentally, atomic bright soliton scattering has only been studied in the regime of negligible interaction energy, where the behavior  resembles the one of non-interacting particles \cite{Dries2010, Marchant2013, Marchant2016}. 


In this paper, we report the observation of nonlinear scattering of an atomic bright soliton in the regime where the interaction energy is of the order of the center of mass kinetic energy \cite{Lepoutre2016}. As the experiment is repeated, we find that the atoms tend to be collectively either reflected or transmitted. More precisely, the histogram of the reflected fraction shows two distinct peaks at low (close to 0) and high (close to 1) reflected fractions, in contrast with the observed bell shaped histogram for non-interacting particles in the single scattering regime. This behavior is a signature of the non-linear behavior of solitons in scattering. We find that Gross Pitaevskii simulations are sufficient to account for our observed double peaked histogram, because of their strong sensitivity to small fluctuations of the experimental parameters and in particular to the soliton velocity. We argue however, that mesoscopic quantum superpositions of all atoms being reflected {\it and} all atoms being transmitted \cite{Weiss2009, Streltsov2009, Streltsov2009b}, could be observable in similar conditions provided that the number of atoms is significantly reduced.

\section{Methods}
Our experiment starts with a  $^{39}$K condensate in the $|F=1, m_F=-1\rangle$ state, produced by evaporative cooling in an optical trap \cite{Salomon2014} close to the 561\,G Feshbach resonance \cite{Derrico2007}. A soliton, containing $N=$5500(800) atoms, is then created by ramping the magnetic field close to the scattering length zero crossing at 504.4\,G \cite{Derrico2007, Lepoutre2016}.  The atoms then have a negative mean-field interaction energy, which binds them together. The elongated trap is made of two horizontal far-detuned optical beams (at 1064\,nm and 1550\,nm),  and it has identical radial frequencies of $ \omega_\perp/2 \pi=195$\,Hz and a longitudinal frequency of $\omega_z/2 \pi=44$\,Hz. 

\begin{figure}[htbp]
\centering
\includegraphics[width=0.45\textwidth]{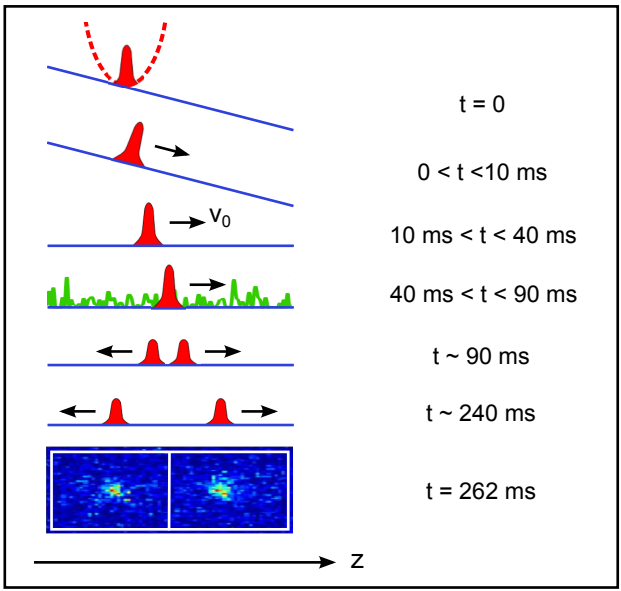}
\caption{(Color online) Schematic of the experimental sequence. A soliton is launched into a 1D waveguide along $z$ (continuous blue line) from a longitudinal trap (dotted red line). The soliton is first accelerated to a controlled velocity $v_0$ before a 1D speckle at 532\,nm (green curve) is shone on the atoms for 50\,ms. The reflected and transmitted parts are finally separated and observed after an additional 150\,ms wait time, when an image of the density distribution is taken.}\label{sequence}
\end{figure}
The soliton scattering in a 1D disordered potential is studied through the measurement of the reflected fraction of the cloud sent with a low velocity in a far off resonance speckle field. The sequence is the following (see Fig.\,\ref{sequence}). The longitudinal (along $z$) confinement is suddenly removed and the soliton starts to propagate along $z$ in a 1D tube. We control the initial longitudinal acceleration  through the addition of a small magnetic field gradient. The latter is subsequently ramped down between 10\,ms and 40\,ms after trap release such that the acceleration then vanishes (see footnote\footnote{In practice, we have an additional residual anti-trapping curvature of frequency $i\times 1.9\,$Hz \cite{Lepoutre2016}, which only plays a role on long time scales and that we take into account in the analysis.}). We choose the initial acceleration in order to reach a velocity of either $v_0=0.51(16)\,$mm.s$^{-1}$ or $v_0=0.90(20)\,$mm.s$^{-1}$, corresponding to a center-of-mass kinetic energy per particle $E_\mathrm{kin}/h=m v_0^2/2h=$13(8)\,Hz or 40(17)\,Hz, where $m$ is the atomic mass and $h$ the Planck constant. The fluctuations of the initial velocity exceed,  by a factor $\sim 25$, those associated with the quantum fluctuations of the soliton center of mass in the ground state of the initial trap. They are due to uncontrolled and undamped residual dipole oscillations in the initial trap.

A 1D disorder potential is then turned on for 50\,ms and the atoms are partially scattered or reflected, since we are in a 1D situation. After a waiting time of 150 ms, the transmitted and reflected components are well separated, and the radial trap is switched off. Each cloud expands for another 22 ms, and the separated components are observed (fig.\,\ref{sequence}) by resonant fluorescence imaging as presented in \cite{Lepoutre2016}. The atom numbers in each component are directly obtained (within a multiplying constant) by integration over two zones corresponding to positive and negative velocities (see Fig.\,\ref{sequence}), whereas the background is estimated from neighboring zones. We thus have a measurement that is independent of any assumption on the cloud shapes. The accuracy of atom number detection permits us to determine the reflected fraction with a 10$\%$ accuracy for each individual run. 

The disorder is created from a laser speckle at 532\,nm, which yields a repulsive conservative potential for the atoms \cite{Clement2006}.  The laser beam, propagating perpendicularly to $z$, passes through a diffusing plate and is focused on the atoms. Its cross-section intensity distribution on the diffusing plate is elliptical, with long axis along $z$ and short axis perpendicular to $z$. The speckle pattern shone on the atoms has an intensity autocorrelation function whose widths along these two directions are respectively $\sigma_z=0.38$\,$\mu$m, and 2.4\,$\mu$m (half-width at $1/\sqrt{\mathrm{e}}$). Along the propagation axis of the laser beam, this autocorrelation width is 10\,$\mu$m. The two correlation lengths perpendicular to $z$  exceed the r.m.s. radial size of the cloud given by the ground state extension of the harmonic oscillator $\sqrt{h/4\pi m \omega_\perp}=0.8\,\mu$m. The disordered potential is thus one-dimensional for the atoms moving along $z$. The disorder correlation width  $\sigma_z=0.38$\,$\mu$m corresponds to  $k \sigma_z=0.12(4)$ and $k \sigma_z=0.21(5)$, where $k=2 \pi m v_0/h$ is the $k$-vector of the de Broglie wave of an individual atom moving at the velocity $v_0$ of our two sets of data. Consequently, individual atoms experience quantum scattering  (quantum tunneling and quantum reflection) in this disorder \cite{Sanchez-Palencia2007, Billy2008}. Scattering experiments with non-interacting atoms at various velocities and disorder amplitudes allow us to calibrate the speckle amplitude (see supplementary material). For the study reported in this paper, we use $V_R/h=13.5(2.0)\,$Hz, where $V_R$ is the mean value of the exponential probability distribution of the potential due to the laser speckle ($V_R$ is equal to both its average and  r.m.s. value). The probability for a single atom to be reflected during its interaction with the speckle is typically $\sim$35\% and we work in the single scattering regime.

\section{Results}
The measurements of the reflected fractions are performed for solitons and for non-interacting clouds. For a scattering length $a=-2.0(2)\,a_0$ ($a_0$ is the Bohr radius), a strongly bound soliton is formed, close to the collapse threshold \cite{Carr2002, Lepoutre2016}. An approximate value of the chemical potential can be obtained based on the 1D formula: $\mu_\mathrm{1D}/h=-\frac{1}{2}m \omega_\perp^2 N^2 a^2/h=-25(12)\,$Hz. This value is comparable to the center of mass kinetic energy per particle and we expect an effect of the interactions in the scattering process. On the contrary, for $a=-0.9(2)\,a_0$ , the interaction energy is barely sufficient to hold the atoms together after the trap release and the cloud is observed to slowly spread at long time. Choosing such a value of the scattering length permits to avoid the spread in velocity that would be given by $\sqrt{h \omega_z/4 \pi m}=0.47\,$mm.s$^{-1}$ for a non-interacting condensate.
Regarding the scattering in the disorder, in this case, the atoms can be considered as non-interacting. 

 \begin{figure}[tbp]
\centering
\includegraphics[width=0.49\textwidth]{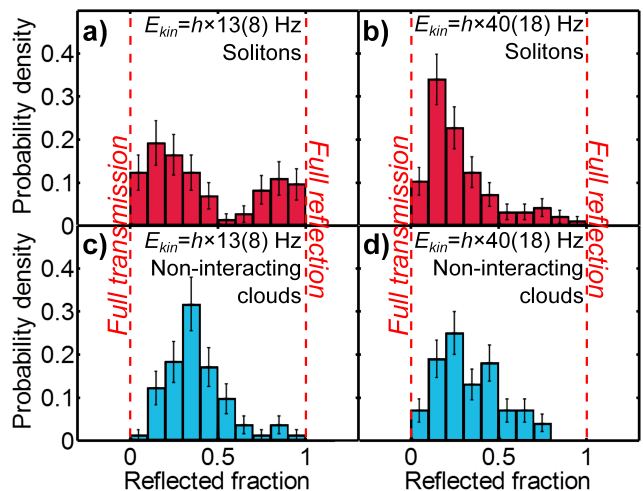}
\caption{(Color online) Histograms of the experimentally measured reflected fractions of solitons (a) and b) in red) and non interacting atoms (c) and d) in blue).  The double peak structure in a) is a clear signature of nonlinear scattering. The chemical potentials of solitons in a) and b) are estimated to $\mu_\mathrm{1D}/h=-25(12)\,$Hz. The center of mass kinetic energies are $E_\mathrm{kin}/h=13(8)$\,Hz in a) and c) and $E_\mathrm{kin}/h=40(17)$\,Hz in b) and d). The error bars are given by $\sqrt{N_\mathrm{b}}$, where $N_\mathrm{b}$ is the number of events in each bin.  The number of repetitions per histogram is $\sim$90.
} \label{hist_exp}
\end{figure}

\begin{figure}[tbp]
\centering
\includegraphics[width=0.49\textwidth]{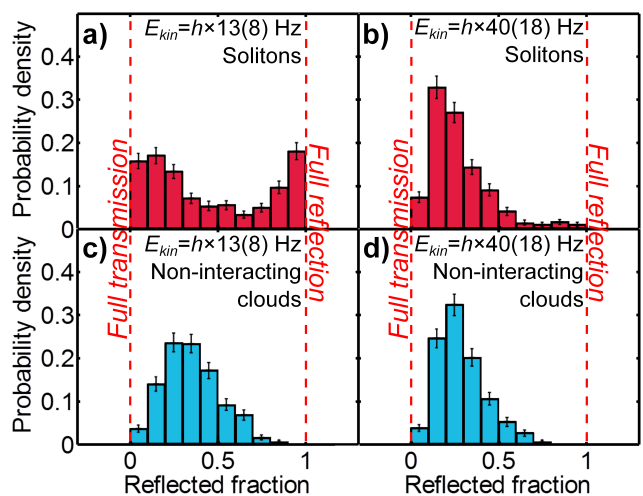}
\caption{(Color online) Histograms of the reflected fractions simulated from the 1D Gross-Pitaeskii equation for our parameters and for random disorders. The chemical potential of solitons a) and b) is adjusted to match the experimental findings to $\mu=-h\times35\,$Hz. The center of mass kinetic energies are $E_\mathrm{kin}/h=13(8)$\,Hz in a) and c) and $E_\mathrm{kin}/h=40(17)$\,Hz in b) and d).  The error bars are given by $\sqrt{N_\mathrm{b}}$, where $N_\mathrm{b}$ is the number of events in each bin. The number of repetitions per histogram is $\sim$ 500.} \label{hist_theo}
\end{figure}

For each set of parameters, we repeat the scattering experiment several times. In similar experimental conditions, the measured reflected fractions fluctuate between 0 and 100\%, as reported in the histograms of the  reflected fractions (Fig.\,\ref{hist_exp}). At $E_\mathrm{kin}/h=13(8)$\,Hz (Fig.\,2a), the histogram shows two distinct peaks centered around reflected fractions of $\sim$0.2 and $\sim$0.85. Moreover, the soliton rarely splits into two equal reflected and transmitted parts. This histogram thus shows a tendency for the atoms to be collectively either reflected or transmitted. This is in contrast with the observed behavior for non-interacting clouds at the same kinetic energy (Fig.\,2c): the histogram then exhibits a single broad peak around a reflected fraction of $\sim 0.35$. This observed striking difference between interacting and non-interacting situations is a clear indication of an effect of the nonlinearity in the scattering of bright solitons.

We now compare those findings with experiments performed at a larger center of mass kinetic energy $E_\mathrm{kin}/h=40(17)$\,Hz (Fig.\,2b and Fig.\,2d). We find that the double peak feature in the histogram obtained with solitons tends to disappear (Fig.\,2b). These additional results show that the ratio $\alpha=-\mu/E_\mathrm{kin}$ is an important parameter, comparing the chemical potential to the kinetic energy.  Its value is respectively $\alpha\sim 2$ and $\alpha\sim 0.6  $ in Fig.\,2a and 2b. In our experiment, the nonlinear behavior is thus observed to set in for $\alpha$ of the order of 1. Note that  when $\alpha >4$, it becomes energetically forbidden to split the soliton in two equal parts \cite{Gertjerenken2012, Helm2014}.

In order to  interpret our results more quantitatively, we compare them with numerical simulations  of the 1D Gross-Pitaevskii equation. For each given set of parameters we find a unique value of the reflected fraction, and in order to compare to our histograms, we repeat the simulations taking into account the fluctuations in velocities and speckle amplitudes corresponding to the ones in the experiments.  Moreover, we also sample over different speckle realizations, although we keep the same speckle pattern in the experiment (see the discussion below).  The simulated histograms (see fig.\,\ref{hist_theo}) are similar to the experimental ones for non-interacting atoms and for solitons with a chemical potential $\mu/h=-35$\,Hz. A good match with the experimental data is obtained in the range $-27\,$Hz$>\mu/h>-$43\,Hz.  Such a chemical potential is in agreement with the previously estimated experimental value. For more negative values of the chemical potential, the simulation results tend toward full reflection or transmission of the solitons. For less negative values of the chemical potential, the results are close to those expected for non interacting atoms, consistently with the importance of the ratio $\alpha=-\mu/E_\mathrm{kin}$. 

One may question the validity of the above comparison, since, experimentally, we do not move the diffusive plate and thus do not change the speckle realization. In fact, the fluctuations in the initial velocity of the condensate lead also to fluctuations in the region of the disorder explored by the atoms, during the period when the disorder is turned on (Fig.\,1). We have checked that simulations with variations in the initial velocity and a fixed typical disorder yield a distribution of the reflected fractions similar to the one obtained with different disorders. Moreover, after tens of repetitions of the experimental cycle, thermal drifts of the position of our trapping beam relative to the speckle would correspond to an additional disorder averaging. We conclude that the Gross-Pitaevskii equation is sufficient to simulate our experimental results, provided that we take into account fluctuations of the experimental conditions. 

It is nevertheless interesting to consider the possibility that the shot to shot variations of the observed reflection coefficient would stem from a mesoscopic quantum superposition of most atoms reflected and most atoms transmitted. Such a beyond mean-field behavior has been theoretically predicted in the case of a quantum reflection of a soliton on a thin barrier, when it is energetically protected from splitting \cite{Weiss2009, Streltsov2009, Streltsov2009b}. In this case, a key parameter is $Nk\sigma_z$ (where $N k$ is the soliton $k$-vector), which governs the scattering of the N-body bound state.  A global quantum behavior is expected only for $Nk\sigma_z$ of the order of 1  or below, or equivalently when the de Broglie wavelength of the giant particle is larger than the defect sizes. With our parameters, $Nk\sigma_z \approx 10^3 \gg 1$, the soliton as a whole is expected to behave classically, with either full transmission or full reflection, depending on the relative value of its kinetic energy compared with the highest potential peak in the explored disorder sample. It rules out an interpretation of our results in terms of mesoscopic quantum superpositions, which should be observable for lower atom numbers.

\section{Conclusion}
In conclusion, we have studied the scattering of bright atomic solitons in a regime where the interaction energy exceeds the center of mass kinetic energy, and compared it to the scattering of non-interacting atoms with the same velocity. We identify a nonlinear regime of scattering that is characterized by a tendency for the soliton to be either fully transmitted or reflected, as clearly visible in the histograms of reflected fractions. This behavior is captured in the Gross Pitaevskii mean-field approach, provided that we take into account the strong sensitivity of the nonlinear behavior to the  fluctuations of the experimental parameters such as the soliton velocity. 

For longer propagation time in the disorder (and possibly slightly higher $\alpha=\mu/E_\mathrm{kin}$), we would enter a multiple scattering regime and should observe the striking situation of a soliton propagating in the disorder without scattering whereas single atoms at the same velocity would be Anderson localized \cite{Kivshar1990, Sanchez-Palencia2007} as previously observed with superfluid helium surface solitons \cite{Hopkins1996}. The soliton is then unaffected by the disorder as a giant classical object. Another interesting possibility would be to replace our static disorder by thermal atoms acting as random moving scatterers. In this case, Brownian motion of the soliton is expected \cite{Aycock2016, McDonald2016}.

Finally, reducing the atom number in the soliton to 10 or 100 particles,  while keeping the same value for the chemical potential \cite{Medley2014}, would permit one to be in the appropriate regime to observe mesoscopic quantum superpositions of the soliton  behaving globally as a giant quantum particle \cite{Streltsov2009}. Such states  would be  interesting for interferometry beyond the standard quantum limit \cite{Jo2007, Veretenov2007, Lee2012, Kasevich2012, Gertjerenken2013, Helm2014, Gertjerenken2015}, and the study of decoherence of these mesoscopic quantum superposition  would be especially interesting. Note also that in this quantum regime, Anderson localization of the whole soliton is  predicted \cite{Sacha2009}.

\acknowledgments
This research has been supported by CNRS, Minist\`ere de l'Enseignement Sup\'erieur et de la Recherche, Direction G\'en\'erale de l'Armement, ANR-12-BS04-0022-01, Labex PALM, ERC senior grant Quantatop, Region Ile-de-France in the framework of DIM Nano-K (IFRAF), EU - H2020 research and innovation program (Grant No. 641122 - QUIC), Triangle de la physique.

\end{document}